\documentclass[
reprint,           
superscriptaddress,
amsmath,           
amssymb,           
aps,               
prd,               
notitlepage,       
floatfix,          
nofootinbib,
]{revtex4-1}

\usepackage{tensor}     
\usepackage{graphicx}   
\usepackage[
colorlinks=true,        
citecolor=blue,         
linkcolor=blue,         
urlcolor=blue           
]{hyperref}             
\usepackage{bm}         
\usepackage{xcolor}     
\usepackage{tabulary}
\usepackage{color}      
\usepackage[utf8]{inputenc} 
\usepackage[section]{placeins} 
\usepackage{url}
\usepackage{blindtext}
\usepackage[T1]{fontenc}
\usepackage[utf8]{inputenc}
\allowdisplaybreaks[4]

\newcommand{\nc}{\newcommand*} 

\nc{\al}{\alpha}
\nc{\s}{\sigma}
\nc{\dt}{\delta}
\nc{\Dt}{\Delta}
\nc{\Ld}{\Lambda}
\nc{\p}{\partial}
\nc{\om}{\omega}
\nc{\Om}{\Omega}
\nc{\rd}{\mathrm{d}}
\nc{\Od}[1]{\mathcal{O}(#1)} 
\nc{\kp}{\kappa}
\nc{\one}{\uppercase\expandafter{\romannumeral1}}
\nc{\two}{\uppercase\expandafter{\romannumeral2}}
\nc{\three}{\uppercase\expandafter{\romannumeral3}}
\def\({\left(}
\def\){\right)}
\def\[{\left[}
\def\]{\right]}
\def\e{\begin{equation}}
\def\q{\end{equation}}
\def\m{\begin{eqnarray}}
\def\n{\end{eqnarray}}
\nc{\Eq}[1]{Eq.~\eqref{#1}}     
\nc{\Fig}[1]{Fig.~\ref{#1}}     
\nc{\Table}[1]{Table~\ref{#1}}  
\nc{\Sec}[1]{Sec.~\ref{#1}}     
\nc{\Msun}{M_\odot}             
\nc{\fpbh}{f_{\mathrm{pbh}}}    
\nc{\fpbhn}{f_{\mathrm{pbh0}}}    
\nc{\mR}{\mathcal{R}} 
\nc{\seq}{\sigma_{\mathrm{eq}}}
\nc{\ogw}{\Omega_{\mathrm{GW}}}
\nc{\gpcyr}{\mathrm{Gpc}^{-3}\,\mathrm{yr}^{-1}}
\nc{\lvc}{LIGO/Virgo} 
\nc{\SNR}{\mathrm{SNR}} 
\nc{\mmin}{{m_{\mathrm{min}}}}
\nc{\mmax}{{m_{\mathrm{max}}}}
\nc{\Mmin}{{M_{\mathrm{min}}}}
\nc{\fmin}{{f_{\mathrm{min}}}}
\nc{\VT}{\mathrm{VT}}
\nc{\rhoGW}{\rho_{\mathrm{GW}}}
\nc{\vth}{\vec{\theta}}
\nc{\vd}{\vec{d}}
\nc{\vla}{\vec{\lambda}}
\nc{\Nobs}{N_{\mathrm{obs}}}
\nc{\av}[1]{\langle #1 \rangle} 
\nc{\km}{\mathrm{km}}
\nc{\Mpc}{\mathrm{Mpc}}
\nc{\Tobs}{T_{\mathrm{obs}}}
\nc{\Ntemp}{N_{\mathrm{temp}}}
\nc{\addref}{[\textcolor{red}{add ref}] } 
\nc{\eg}{\textit{e.g.~}}
\nc{\app}{\approx}
\nc{\hf}{\frac{1}{2}}
\nc{\discuss}{\textcolor{red}{Add discussion here!}}
\nc{\red}[1]{\textcolor{red}{#1}}
\nc{\mH}{\mathcal{H}}
\nc{\cs}{c_s^2}
\nc{\Sij}[1]{S_{ij}^{(#1)}}
\nc{\vi}[1]{v_i^{(#1)}}
\nc{\no}{\nonumber}
\def\<{\left\langle}
\def\>{\right\rangle}

\nc{\bk}{\bm{k}}
\nc{\bq}{\bm{q}}
\nc{\bp}{\bm{p}}
\nc{\bl}{\bm{l}}
\nc{\bx}{\bm{x}}
\nc{\be}{\mathbf{e}}
\nc{\mS}{\mathcal{S}}
\nc{\te}{\tilde{\eta}}
\nc{\tp}{\tilde{p}}
\nc{\tk}{\tilde{k}}
\nc{\tx}{\tilde{x}}
\nc{\tF}{\tilde{F}}
\nc{\tA}{\tilde{A}}
\nc{\mkpq}{|\bk-\bp-\bq|}
\nc{\mpq}{|\bp-\bq|}
\nc{\mkp}{|\bk-\bp|}
\nc{\mSi}[1]{\mS^{(#1)}({\bk, \eta})}
\nc{\vk}{\vec{k}}
\nc{\kstar}{k_*}
\nc{\xstar}{x_*}
\nc{\mpbh}{m_{\rm{pbh}}}

\renewcommand{\vec}[1]{\boldsymbol{#1}} 

\begin{document}
	
\title{The three body first post-Newtonian effects on the secular dynamics of a compact binary near a spinning supermassive black hole}
	
\author{Yun Fang}
\email{fangyun@mail.itp.ac.cn}
\affiliation{CAS Key Laboratory of Theoretical Physics, 
Institute of Theoretical Physics, Chinese Academy of Sciences,
Beijing 100190, China}
\affiliation{School of Physical Sciences, 
University of Chinese Academy of Sciences, 
No. 19A Yuquan Road, Beijing 100049, China}

\author{Qing-Guo Huang}
\email{huangqg@itp.ac.cn}
\affiliation{CAS Key Laboratory of Theoretical Physics, 
Institute of Theoretical Physics, Chinese Academy of Sciences,
Beijing 100190, China}
\affiliation{School of Physical Sciences, 
University of Chinese Academy of Sciences, 
No. 19A Yuquan Road, Beijing 100049, China}
\affiliation{School of Fundamental Physics and Mathematical Sciences
Hangzhou Institute for Advanced Study, UCAS, Hangzhou 310024, China}
\affiliation{Center for Gravitation and Cosmology, 
College of Physical Science and Technology, 
Yangzhou University, Yangzhou 225009, China}
\affiliation{Synergetic Innovation Center for Quantum Effects and Applications, 
Hunan Normal University, Changsha 410081, China}
	
\date{\today}

\begin{abstract}

The binary black holes (BBHs) formed near the supermassive black holes (SMBHs) in the galactic nuclei would undergo eccentricity excitation due to the gravitational perturbations from the SMBH and therefore merger more efficiently. 
In this paper, we study the coupling of the three body 1st post-Newtonian (PN) effects with the spin effects from the SMBH in the hierarchical triple system. 
We extend previous work by including the coupling between the de Sitter precession and the Lense-Thirring precession from the SMBH spin.
This coupling includes both the precessions of the inner orbit angular momentum and the Runge-Lenz vector around the outer orbit angular momentum in a general reference frame. 
We find the change of the (maximal) eccentricity in the neighboring Kozai-Lidov cycles due to spin effects is detectable by LISA  in the future. 
Our general argument on the coupling of the three body 1PN effects in three body systems could be extended to any other situation as long as the outer orbital plane evolves. 

\end{abstract}

\pacs{???}
	
\maketitle

\section{introduction}
	
The first detection of gravitational wave (GW) from a merger event of a binary black hole (BBH) by \lvc  \cite{Abbott:2016blz} in 2015 showed the tremendous success of general relativity (GR) and opened an era of gravitational wave astrophysics. Up to now, the LIGO-VIRGO collaboration observed eleven gravitational wave signals from compact binary mergers during the first and second runs (O1 and O2) \cite{LIGOScientific:2018mvr}, and the third observation run (O3) is undergoing since April 2019 \cite{GWdata}. 
Currently, we have five ground-based detectors \cite{2015CQGra..32g4001L, 2012JInst...7.3012A, Aso:2013eba, Grote_2010} that focus on the merger and ringdown phase of GW sources which is characterized by a frequency of 10Hz to 1000Hz and a strain of order $10^{-22}$.

The space-based detector Laser Interferometer Space Antenna (LISA) is expected to explore the lower frequency GW sources with frequency range from $10^{-4}$ Hz to $1$ Hz and characteristic strain of order $10^{-21}$ \cite{Seoane:2013qna}. 
The  DECi-hertz Interferometer Gravitational wave Observatory (DECIGO) is aiming to fill the gap between LIGO and LISA with frequency band around $10^{-2}$ Hz to $10$ Hz \cite{Sato:2017dkf}. There are several other big projects on the space detectors in the future: advanced LISA (aLISA) \cite{aLISAGW},  TianQin and Taiji in China \cite{Luo:2015ght, Guo:2018npi}. The space and ground based gravitational wave detectors could cover all the inspiral-merger-ringdown phase of compact binaries. The observations of GWs enable us to figure out the binary formation channels (see e.g. \cite{Breivik:2016ddj, Rodriguez:2016vmx}), test the validity of GR in the strong-field regime (see e.g. \cite{Berti:2018cxi, Berti:2018vdi, Wang:2020cub}), and shed light on the gravitational wave astrophysics \cite{Preto_2011, Barausse:2014pra, Barsotti2018, McWilliams:2019fng}. 

The origin of the LIGO/Virgo BBHs is a mystery. Conventionally, they are believed to be formed as the remnant of  massive binary stars or they are formed dynamically in the star clusters \cite{TheLIGOScientific:2016htt}. While according to the recent studies, the centers of galaxies \cite{Miller:2008yw}, especially those hosting supermassive black holes (SMBHs) \cite{Antonini:2012ad} are also important places for BBHs to form. In these environments, the merger rate of BBHs could be enhanced to a significant fraction of the LIGO/Virgo event rate due to the complex astrophysical dynamics \cite{Hong:2015asd, VanLandingham:2016ccd, Hoang:2017fvh, Petrovich:2017otm, Bartos:2016dgn, Stone:2016wzz, McKernan:2017umu, Chen:2017xbi, Sedda:2018znc, Hamers2018ApJ, Fragione:2018yrb, Fragione2019MNRAS, Rasskazov2019ApJ}. And a fraction of the BBHs in galaxy centers could either form at \cite{Inayoshi:2017hgw, Stone:2016wzz, Bartos:2016dgn, McKernan:2017umu, Secunda:2018kar} or be captured to places very close to the SMBHs \cite{Addison:2015bpa, Chen:2018axp}. 

The BBH formed near the SMBH are in a stable hierarchical triple system. Here, we call it the "SMBH- BBH" triple system, where the BBH as the inner binary and their center of mass revolving around the SMBH at a larger outer orbit. The BBH is perturbed by the SMBH, and the dominant Newtonian quadrupole perturbation causes the ``Kozai-Lidov" oscillation \citep{Kozai:1962zz, Lidov:1962zs, Naoz:2016vh} on the BBH orbit.  
The Kozai-Lidov oscillation is described by the exchange between the inner orbital eccentricity and the inclination angle, as a result of the interaction between the inner and the outer orbit angular momentum. 

The general relativistic effects are proved to be important in the secular evolution of three body systems. 
The relativity precession of the inner orbital pericenter is known to suppress the Kozai-Lidov oscillation \cite{Wen:2002km, Naoz:2016vh} if the time scale of the former is shorter than the later.  And the gravitational radiation is known to circularize and shrink the binary orbit \cite{Peters:1963ux, Wen:2002km}. 
These two effects are due to the post-Newtonian (PN) interactions in the inner binary.  
The three body PN effects are coming from the post-Newtonian interactions between the three bodies which are considered in \cite{Naoz:2012bx, Will:2013cza, WillPRL2018, Lim:2020cvm}. 
It is found in \cite{Naoz:2012bx} that there is a resonant eccentricity excitation behavior in the three bodies under PN dynamics in some parameter space by conducting an orbit-averaged three body 1PN Hamiltonian, though, the Hamiltonian approach does not resolve all of the three body PN effects which is stated in \cite{Will:2013cza, Lim:2020cvm}. 
Will points out that to find the full solution to the problem of secular evolution with quadrupole and 1PN effects together, the cross terms in the acceractions \cite{Will:2013cza} and a multiple-scale analysis to account for the corrections of the periodic effects are needed \cite{WillPRL2018}. When the mass of the inner binary is relatively small,  it is found in \cite{Lim:2020cvm}  there are three dominant three body PN effects, where the main effect is the de Sitter precession \cite{de-Sitter1916} which comes directly from the accelerations. 
The above works either consider the three body systems at 1PN order or assume Schwarzschild black holes. In these cases, the outer orbital plane is nearly a constant, and thus the de-Sitter precession is decoupled in the zero order (of multiple-scale analysis) secular evolutions since the equations of motion do not depend on the longitude of ascending nodes.  While the observations indicate that the SMBHs are universally spinning \cite{Reynolds:2013rva, Reynolds:2013qqa}. Our recent studies \cite{Fang:2019hir, Fang_2019} show that when it comes to a three body system where the third body is a spinning SMBH, the Lense-Thirring precession of the outer orbit will cause the Newtonian quadrupole secular dynamics to depend on the angle between the two orbital line of nodes $(\Omega-\Omega_3)$, here we call it the "generalized Kozai-Lidov oscillation", thus lead to different evolutionary behaviors. So, it is worthy to study the dynamical behaviors depend on the angles of nodes, such as the Lense-Thirring precession and the de-Sitter precession.

The de-Sitter precession is previously considered as a sub-leading effect in the three body systems when the third body is of smaller mass \cite{Will:2013cza, WillPRL2014}, or in the case when it is decoupled in the zero order secular equations \cite{Lim:2020cvm}. While the spin effect from the SMBH will cause it to couple in the secular dynamics through the generalized Kozai-Lidov oscillation. And due to the large mass of the SMBH, the de-Sitter precession could reach to or even larger than the amplitude of the binary 1PN effect. 
Liu and Lai in the paper \cite{Liu:2019tqr} noticed that the de-Sitter precession of the inner orbit also becomes important when combined with the Lense-Thirring precession of the outer orbit.  
However, they only added the precession of the inner orbit angular momentum alone the outer orbit one in that work, while we point out in this work is just a part of the de-Sitter precession in this case.  In this work, we study the coupling of the three body 1PN effect which is dominated by the de-Sitter precession in the SMBH-BBH system where the SMBH is spinning, thus revolving this system up to 1.5PN order. We analysis theoretically the condition when the three body 1PN effect is significant in the dynamics of our SMBH-BBH triple system. And we discuss the impact of the de-Sitter precession on the gravitational waves of the BBH by LISA's detection. Further more, we give a proposal to find the unique characteristic left by SMBH spin effect on the BBH waveforms.

This paper is organized as follows. We calculate the equations of motion in section \ref{EOM}. In subsections \ref{full1PN} and \ref{1_5PN}, we analysis the Newtonian quadrupole order, full 1PN order and the 1.5PN order equations of motion. In section \ref{secularevolutions} we calculate systematically the three body 1PN order effects in a general reference frame and discuss the connections between the secular equations of motion listed in section \ref{EOM}. We present our numerical results in section \ref{numericalresult}. In subsection \ref{dynamicalevolution}, we show numerical results of the general relativity effects calculated and considered in this work. And in subsection \ref{onGW}, we show typical characteristics on GW singles due to spin effects. We conclude our paper in section  \ref{conclusion}. 

Throughout this paper we use the natural units with $c=G=1$ in our calculations. 
  
\section{Equations of motion up to 1.5PN order}
\label{EOM}
\subsection{Full 1PN dynamics}
\label{full1PN}
We now consider a hierarchical three-body system in which the binary bodies of mass $m_1$ and $m_2$ are in a close orbit with separation $r$, their center of mass revolving around a SMBH of mass $m_3$ at a much larger distance $R (\gg r)$.   We define the relative separation vector of the binary system and the vector from the center of mass of the binary to the SMBH by
\begin{equation}
{\bm x} \equiv {\bm x}_1 - {\bm x}_2 \,, \quad {\bm X} \equiv {\bm x}_0 - {\bm x}_3  \,,
\end{equation}
where
\begin{equation}
{\bm x}_0 \equiv \frac{m_1 {\bm x}_1 + m_2 {\bm x}_2}{m} \,,
\end{equation}
is the center of mass of the inner binary, and $m \equiv m_1 + m_2$.   
We work in the center of mass-frame of the entire system,  thus,
\begin{equation}
m_1 {\bm x}_1 + m_2 {\bm x}_2 + m_3 {\bm x}_3 = m{\bm x}_0 + m_3 {\bm x}_3 = 0 \,,
\label{centerof mass}
\end{equation}
where we have ignored the post-Newtonian corrections to the center of mass. 
Then, the positions of the three bodies are, 
\begin{equation}
{\bm x}_1 =  \frac{m_2}{m} {\bm x} + \frac{m_3}{M} {\bm X} ,\, 
 {\bm x}_2 = - \frac{m_1}{m} {\bm x} + \frac{m_3}{M} {\bm X} ,\,
 {\bm x}_3 = -\frac{m}{M} {\bm X} ,
 \label{position}
\end{equation}
where $M = m_1 + m_2 + m_3$ is the total mass. Since the mass of SMBH is much larger than the binary system, with $m_3\gg m$, the center of mass frame of the entire system is set to the position of $m_3$. Thus Eq.~(\ref{position}) is simplified to
\begin{equation}
{\bm x}_1 =  \frac{m_2}{m} {\bm x} + {\bm X} ,\,
{\bm x}_2 = - \frac{m_1}{m} {\bm x} + {\bm X} ,\,
{\bm x}_3=0 ,
\end{equation}
We also define the velocities ${\bm v} \equiv d{\bm x}/dt$, ${\bm V} \equiv d{\bm X}/dt$, accelerations ${\bm a} \equiv d{\bm v}/dt$, ${\bm A} \equiv d{\bm V}/dt$, distances $r \equiv |{\bm x}|$, $R \equiv |{\bm X}|$, and unit vectors ${\bm n} \equiv {\bm x}/r$, ${\bm N} \equiv {\bm X}/R$. 

The accelerations are directly computed with the post-Newtonian N-body equations of motion, which is commonly referred to as the Einstein-Infeld-Hoffman  equations of motion \cite{Einstein1938}:

\begin{widetext}  
\begin{eqnarray} \label{eq:EIH}
{\bm a}_a &=&  -\sum_{b \ne a} \frac{m_b {\bm x}_{ab}}{r_{ab}^3}
\nonumber \\
&&  +  \sum_{b \ne a} \frac{m_b {\bm x}_{ab}}{r_{ab}^3}
\biggl [ 4 \frac{m_b}{r_{ab}} + 5\frac{m_a}{r_{ab}} +
\sum_{c \ne a,b} \frac{m_c}{r_{bc}} 
+ 4 \sum_{c \ne a,b} \frac{m_c}{r_{ac}}
\nonumber \\
&& 
- \frac{1}{2} \sum_{c \ne a,b} \frac{m_c}{r_{bc}^3} (\bm{x}_{ab} \cdot \bm{x}_{bc}) - v_a^2 + 4 \bm{v}_a \cdot \bm{v}_b - 2 \bm{v}_b^2 + \frac{3}{2} (\bm{v}_b \cdot \bm{n}_{ab} )^2 \biggr ]
\nonumber \\
&& 
-\frac{7}{2}   \sum_{b \ne a} \frac{m_b}{r_{ab}} \sum_{c \ne a,b} \frac{m_c \bm{x}_{bc}}{r_{bc}^3}  
+  \sum_{b \ne a} \frac{m_b}{r_{ab}^3}  {\bm x}_{ab} \cdot (4 \bm{v}_a - 3 \bm{v}_b ) (\bm{v}_a - \bm{v}_b ) \,,
\end{eqnarray}
where $r_{ab}=|{\bm x}_{ab}|=|{\bm x}_a-{\bm x}_b|$, $\bm{n}_{ab} = \bm{x}_{ab}/r_{ab}$, ${\bm v}_a=d {\bm x}_a/dt$, and $a$, $b$, $c$ denotes 1, 2, 3. 

The inner binary acceleration could be decomposed as follows,
\m  \label{acce}
{\bm a} &=& - \frac{ m{\bm n}}{r^2} - \frac{m_3 \,r}{R^3} \left [ {\bm n} -3({\bm n} \cdot {\bm N}) {\bm N} \right ]
+ [{\bm a}]_{\rm binary 1PN} 
\nonumber \\
&& + [{\bm a}]_{\rm 3body 1PN} + O\left (\frac{ m m_3 }{ R^3}\right )  +O\left (\frac{ {m_3}^2 r}{ R^4}\right ) 
+ O\left (\frac{ {m}^2 r}{ r^3 R}\right )  + ... \,,
\n
where we have expanded the Newtonian perturbation term from the third body to quadrupole order. And $[{\bm a}]_{\rm binary 1PN}$ is the 1PN order acceleration of the binary system ($m_1$ and $m_2$), while $[{\bm a}]_{\rm 3body 1PN}$ is the 1PN order acceleration contribute by the three body interactions. To leading order, they are 
\m   \label{abinary}
[{\bm a}]_{\rm binary 1PN} &=& \frac{m{\bm n}}{r^2} \left [(4+2\eta) \frac{m}{r} - (1+3\eta)v^2 + \frac{3}{2} \eta ({\bm n}\cdot {\bm v})^2 \right ] + (4-2\eta) \frac{m({\bm n}\cdot {\bm v}) {\bm v}}{r^2} \,,
\n

\m  \label{a3body1PN}
[{\bm a}]_{\rm 3body 1PN} &=& \frac{5  m m_3 {\bm n}}{r^2 R}  + \frac{ m}{r^2} \{ [ \frac{3}{2} ({\bm n}\cdot{\bm V})^2-2 \Delta\ {\bm v} \cdot {\bm V}+ V^2 ] {\bm n} - \Delta ({\bm n} \cdot {\bm V}) {\bm v} \}  \nonumber \\
   &&+ \frac{ m_3}{R^2} \[4 {\bm v} \cdot {\bm N} ({\bm V}-\Delta\  {\bm v})+ (\Delta\  v^2 -2 {\bm v}\cdot{\bm V}) {\bm N} +4 ({\bm V}\cdot{\bm N})  {\bm v} \]  \nonumber \\
     &&+ \frac{\Delta  m m_3}{2 r R^2} [9 ({\bm n} \cdot {\bm N}) {\bm n} - {\bm N}] , 
\n
where $\eta={m_1 m_2 \over m^2}$, $\Delta={m_1-m_2 \over m}$. We only keep the dominant terms in $[{\bm a}]_{\rm 3body 1PN}$ which are combined of ${m\over r^2}$ or ${m_3\over R^2}$ with $v^2(\sim{m\over r})$, $V^2(\sim{m_3\over R})$, or ${\bm v}\cdot {\bm V}$. We drop the sub-leading terms in $[{\bm a}]_{\rm 3body 1PN}$ since here we are considering the dominant effect regarding the three-body 1PN interactions which could be the same order or even larger than the binary 1PN effect due the large mass of the SMBH. 

We treat the acceleration of the outer binary as the similar way: 
\m  \label{Acce}
{\bm A} &=& - \frac{M{\bm N}}{R^2} + \frac{3}{2}\frac{M \eta r^2}{R^4} \left [ {\bm N} \left (1- 5({\bm n} \cdot {\bm N})^2 \right) + 2{\bm n} ({\bm n} \cdot {\bm N})\right ]
+ [{\bm A}]_{\rm binary 1PN}  \nonumber \\
&&+ [{\bm A}]_{\rm 3body 1PN} + O\left (\frac{ m m_3 }{ R^3}\right )  +O\left (\frac{ {m_3}^2 r}{ R^4}\right ) 
+ O\left (\frac{ {m}^2 r}{ r^3 R}\right )  + ... \,,
\n
where  
\m
[{\bm A}]_{\rm binary 1PN}&=& \frac{ m_3 {\bm N}}{R^2} (\frac{4  m_3}{R}-V^2)+\frac{4  m_3}{R^2} ({\bm V} \cdot {\bm N}) {\bm V} , 
\label{Abinary}
\n
\m   \label{A3body1PN}
[{\bm A}]_{\rm 3body 1PN} &=& \frac{\eta   m {\bm n}}{r^2} \{ \frac{\Delta\ m }{r}  - 3 ({\bm n} \cdot {\bm v}) ({\bm n} \cdot {\bm V})+\Delta  [ \frac{3}{2} ({\bm n} \cdot {\bm v})^2- v^2 ] +2 {\bm v} \cdot {\bm V} \}  \nonumber \\
&& + \frac{\eta  m {\bm v}}{r^2} (2\ {\bm n} \cdot {\bm V}-\Delta\  {\bm n} \cdot {\bm v}) + \frac{\eta  m_3}{R^2} [ 4 ( {\bm v} \cdot {\bm N}) {\bm v} - v^2 \ {\bm N} ]   \nonumber \\
&&+ \frac{\eta  m m_3}{r R^2} [ {\bm N}-4( {\bm n} \cdot {\bm N}){\bm n} ]  , 
\n
\end{widetext} 
The first terms in Eq.~(\ref{acce}) and (\ref{Acce}) are the leading Newtonian gravitational force which form the Kepler orbit of the inner and outer binary, and the rests in Eqs.~(\ref{acce}) and (\ref{Acce}) are all treated as perturbations. The second terms are the Newtonian quadrupole forces which cause the Kozai-Lidov oscillation (see e.g. \cite{Will:2017vjc}). The binary 1PN acceleration contains the standard terms for a body in orbit around a point mass $m$ (or $m_3$). The three body 1PN accelerations come from the leading three body interactions at 1PN which result the de-Sitter precession as is presented in the section \ref{secularevolutions}. 

\subsection{1.5 PN dynamics from the spin of the SMBH}
\label{1_5PN}
In the previous paper \cite{Fang:2019hir, Fang_2019}, we studied the spin effects from the SMBH on the dynamical evolution of a nearby BBH. In the SMBH-BBH triple system, the gravitational potential is dominated by the mass of the SMBH which contributes the electrical part of dynamics. And similarly, for a relatively large spin parameter of the SMBH, its spin angular momentum will dominant the total angular momentum of this system which contributes the magnetical part of the dynamics \cite{Nichols:2011pu, Fang:2019hir, Fang_2019}. The spin of the SMBH will induce a strong gravitomagnetic field (denoted as ${\bm {\mathcal{H}}}$) in its spacetime. The BBH moving close to the rotating SMBH will feel the gravitomagnetic froce ${\bm v}_a\times {\bm {\mathcal{H}}}$ \cite{Nichols:2011pu, Thorne:1984mz}, and the field ${\bm {\mathcal{H}}}$ is related to the spin momentum ${\bm S}$ as 
\m
{\bm{\mathcal{H}}}={\bm \nabla}\times (-2{ {\bm{S}} \times {\bm r} \over r^3}), 
\n
where ${\bf{S}}=a m_3 {\bf{s}}$, $a/m_3$ is the dimensionless spin parameter, ${\bf s}$ is the spin direction vector, and ${\bm r}$ here denotes the position of the binary relative to the SMBH. \\

Decomposing the gravitomagnetic force into the inner and outer orbit equation of motion, the accelerations are dominated by \cite{Fang:2019hir}
\m 
\mathbf{a}_{[1.5\text{PN,spin}]}&\simeq&2 a m_3\mathbf{v}\times {(\mathbf{e}_Z-3{(\mathbf{e}_Z \cdot \mathbf{N})\mathbf{N}})\over R^3},
\label{inner_spin_acceleration}\\
 \mathbf{A}_{[1.5\text{PN,spin}]}&\simeq&2 a m_3\mathbf{V}\times {(\mathbf{e}_Z-3{(\mathbf{e}_Z \cdot \mathbf{N})\mathbf{N}})\over R^3}, 
\label{outer_spin_acceleration}
\n 
The force in~(\ref{outer_spin_acceleration}) causes the Lense-Thirring precession of the out orbit while the force in~(\ref{inner_spin_acceleration}) causes another precession on the inner orbit which is listed in the next section.  

\section{Secular evolution of the orbit elements}
\label{secularevolutions}
\begin{figure}[tp] 
\centering
\includegraphics[width=1.05 \linewidth]{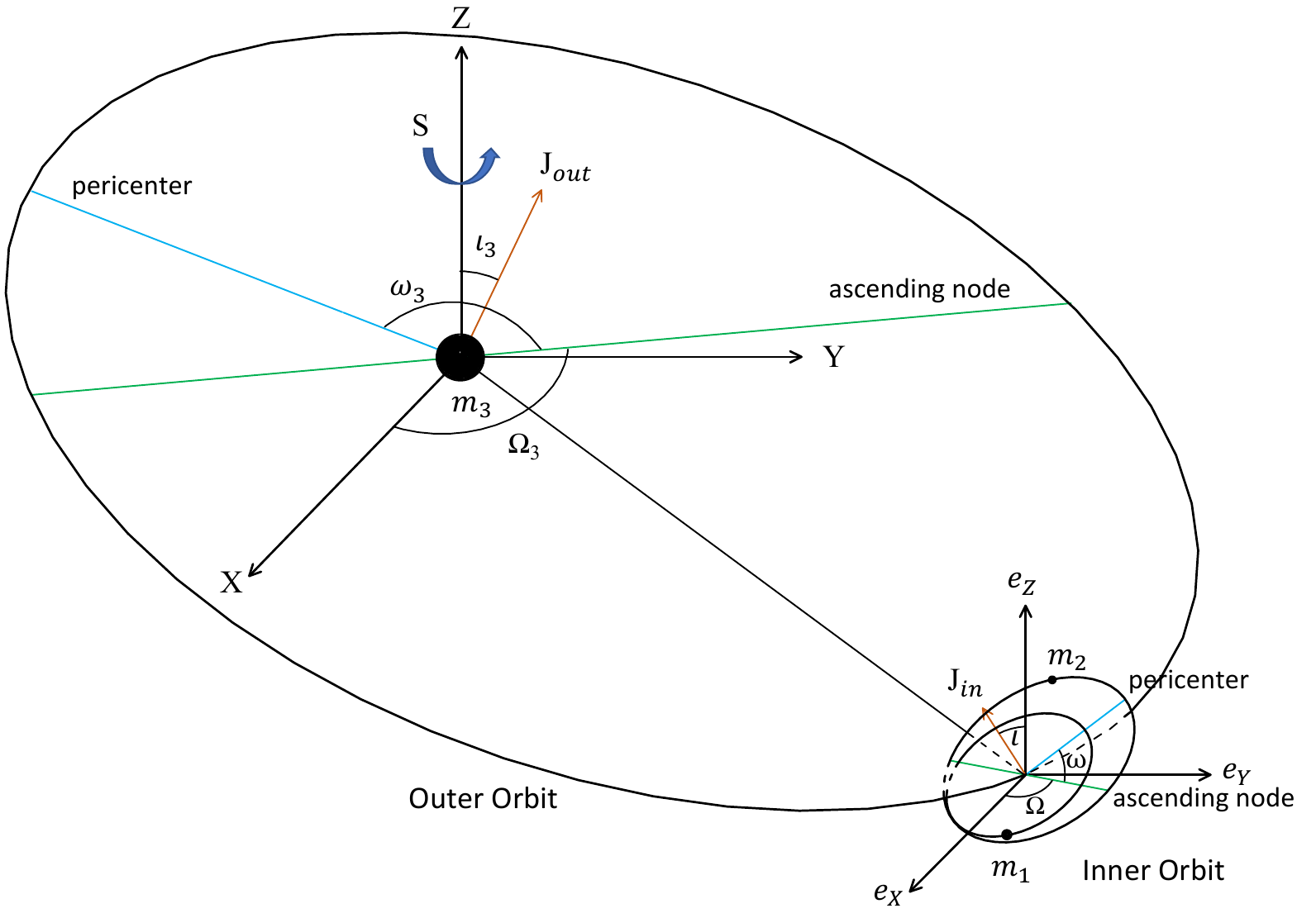} 
\caption{Orbits and angles see from a fixed reference frame where the $Z$ axis is in the direction of spin}
\label{orbits}
\end{figure}

We are interested in the secular evolutions which are left after a complete evolution of the inner and outer orbit. This is obtained by average the Lagrange planetary equations over the period of the inner and outer orbital (see e.g. \cite{Poisson:2014aa}).  
We denote the inner and outer orbits with the time-dependent osculating orbital elements $\{p, e, \omega, \Omega, \iota\}$ and $\{P, E, \omega_3, \Omega_3, \iota_3\}$ respectively. See Fig.~\ref{orbits} for details. The positions and velocities of each orbit are defined in terms of the orbital elements as  
\m
    \bm{r} & = & p \mathbf{n}/ ( 1+e \cos \phi ), \nonumber \\
    \bm{v} & = & \sqrt{\frac{ m}{p}}\Big \lbrack e \sin \phi \ \mathbf{n} + ( 1 + e \cos \phi ) \bm{\lambda}\Big \rbrack, \nonumber \\
    \bm{R} & = & P \mathbf{N} / ( 1+E \cos \Phi ), \nonumber \\
    \bm{V} & = & \sqrt{\frac{ M}{P}}\Big \lbrack E \sin \Phi \  \mathbf{N} + ( 1 + E \cos \Phi ) \bm{\Lambda}\Big \rbrack, 
\n
where the bases $(\mathbf{n},\bm{\lambda},{\mathbf{\hat{h}}})$ and $(\mathbf{N},\mathbf{\Lambda},\mathbf{H})$ are defined on the inner and outer orbital plane which are related to the reference frame $(\mathbf{e}_X, \mathbf{e}_Y, \mathbf{e}_Z)$ by Euler angles \cite{Poisson:2014aa}: 
\m
\mathbf{n}&=&[\cos{\Omega}\cos{(\omega+\phi)}-\cos{\iota}\sin{\Omega}\sin{(\omega+\phi)}]\mathbf{e}_X  \nonumber \\
&&+[\sin{\Omega}\cos{(\omega+\phi)}+\cos{\iota}\cos{\Omega}\sin{(\omega+\phi)}]\mathbf{e}_Y \nonumber  \nonumber \\
&&+\sin{\iota}\sin{(\omega+\phi)}\mathbf{e}_Z,    \nonumber \\
\bm{\lambda}&=&{d\mathbf{n}\over d\phi},\ {\mathbf{\hat{h}}}=\mathbf{n}\times\bm{\lambda}, \\
\mathbf{N}&=&[\cos{\Omega_3}\cos{(\omega_3+\Phi)}-\cos{\iota_3}\sin{\Omega_3}\sin{(\omega_3+\Phi)}]\mathbf{e}_X \nonumber\\
&&+[\sin{\Omega_3}\cos{(\omega_3+\Phi)}+\cos{\iota_3}\cos{\Omega_3}\sin{(\omega_3+\Phi)}]\mathbf{e}_Y \nonumber\\
&&+\sin{\iota_3}\sin{(\omega_3+\Phi)}\mathbf{e}_Z, \nonumber\\
\bm{\Lambda}&=&{d\mathbf{N}\over d\Phi},\ {\mathbf{H}}=\mathbf{N}\times\bm{\Lambda}.  
\n
And the semi-major axis for the inner and outer orbits are respectively $\alpha=p/(1-e^2)$ and $\mathcal{A}=P/(1-E^2)$. \\

We difine the perturbing accelerations as $\delta \mathbf{a}=\mathbf{a}+{m\over r^2}\mathbf{n}$ and $\delta \mathbf{A}=\mathbf{A}+{m_3\over R^2}\mathbf{N}$. 
The orbits are perturbed from Kepler orbit. Take the the inner orbit for example,  
the equation of motion is govern by 
\e
{d \bm{\mathit{h}} \over dt}=\bm{r}\times {\delta \mathbf{a}}, \ m{d\mathbf{Q}\over dt}=\delta \mathbf{a} \times \bm{\mathit{h}} +\mathbf{v}\times (\bm{r}\times \delta \mathbf{a}), 
\label{EOM_vector}
\q
where $ \bm{\mathit{h}} \equiv \bm{r}\times \mathbf{v}=\sqrt{m p} {\mathbf{\hat{h}}}$, and $\mathbf{Q}$ is the Runge-Lenz vector which is defined by $\mathbf{Q}\equiv {\mathbf{v}\times \bm{\mathit{h}} / m}-\mathbf{n}=e (\cos{\phi} \mathbf{n}-\sin{\phi}\bm{\lambda})$. 

The equations of motion of orbital elements are obtained by resolving the equations in (\ref{EOM_vector}) as,  
\begin{align}  \label{innerelements}
{dp\over dt}=&2\sqrt{p^3\over m}{\mathcal{S}\over 1+e \cos{\phi}}, 
 \nonumber \\
{de\over dt}=&\sqrt{p\over m}(\sin{\phi}~\mathcal{R}+{2\cos{\phi}+e+e{\cos^2{\phi}}\over 1+e \cos{\phi}}\mathcal{S}), 
 \nonumber \\
{d{\varpi} \over dt}=&{1\over e}\sqrt{p\over m}(-\cos{\phi}~\mathcal{R}+{2+e\cos{\phi}\over 1+e \cos{\phi}}\sin{\phi}~\mathcal{S}), 
 \nonumber \\
{d\iota\over dt}=&\sqrt{p\over m}{\cos{(\omega+\phi)}\over 1+e \cos{\phi}}\mathcal{W}, 
 \nonumber \\
\sin{\iota}{d\Omega\over dt}=&\sqrt{p\over m}{\sin{(\omega+\phi)}\over 1+e \cos{\phi}}\mathcal{W}, 
\end{align}
where $\mathcal{R}=\mathbf{n}\cdot\delta\mathbf{a}$, $\mathcal{S}=\bm{\lambda}\cdot \delta\mathbf{a}$, $\mathcal{W}={\mathbf{\hat{h}}} \cdot \delta\mathbf{a}$, and $\dot{\omega}$ could be obtained by $\dot{\omega}=\dot{\varpi}-\dot{\Omega}\cos{\iota}$. 
The evolution of outer orbital elements could be obtained analogously by replacing all the elements in the inner orbit to the outer orbital ones, specifically, by repalcing $e \to E, p \to P, m \to M,\phi \to \Phi$, $\iota \to \iota_3$, $\Omega\to \Omega_3$, $\omega\to \omega_3$, and $\mathcal{R}_3=\mathbf{N}\cdot\delta\mathbf{A}$, $\mathcal{S}_3=\mathbf{\Lambda}\cdot \delta\mathbf{A}$, $\mathcal{W}_3={\mathbf{H}}\cdot \delta\mathbf{A}$. 

The secular evolution of the orbital elements are calculated with a double-orbit-average as following
\m
\langle \mathcal{F} \rangle= {1\over T_{\text{out}}}{1\over T_{\text{in}}}\int_{0}^{T_{\text{out}}}\int_{0}^{T_{\text{in}}} \mathcal{F} dt dt', 
\label{averaget}
\n
where $\mathcal{F}$ denote the all the elements in the left of Eq.~(\ref{innerelements}), $T_{\text{in}}$ and $T_{\text{out}}$ are the orbital periods. For convenience of calculation, we change the integration on time to that on the true anomaly $\phi$ and $\Phi$, by $dt=\sqrt{p^3/ m}(1+e\cos{\phi})^{-2}d\phi$ and $dt'=\sqrt{P^3/ m_3}(1+E\cos{\Phi})^{-2}d\Phi$. Thus the average in Eq.~(\ref{averaget}) becomes 
\begin{align}
\langle \mathcal{F} \rangle=& {1 \over 4 \pi^2} (1-e^2)^{3/2} (1-E^2)^{3/2}
 \nonumber \\
&\int_{0}^{2\pi}\int_{0}^{2\pi}{ \mathcal{F}  \over (1+e\cos{\phi})^{2} (1+E\cos{\Phi})^{2} }d\phi d\Phi, 
\end{align}

The Newtonian quadrupole perturbing accelerations in Eqs.~(\ref{acce}) and (\ref{Acce}) result to the Kozai-Lidov formula in the most general form  as follows \cite{Fang:2019hir}, 
\begin{widetext}
\medskip
\noindent
{\bf Quadrupole order}
\begin{align} \label{quadrupolegeneral}
{de \over d\tau}&={15 \pi  \alpha ^3 e \sqrt{1-e^2} m_3 \over 16 \mathcal{A}^3 (1-{E}^2)^{3/2} {m} } \biggl [ \sin ^2\iota _3 (\cos 2 \iota+3) \sin 2 \omega \cos (2 \Omega -2 \Omega_3)+4 \sin ^2\iota _3 \cos \iota  \cos 2 \omega \sin (2 \Omega -2 \Omega _3) 
 \nonumber \\
&-4 \sin 2 \iota _3 \sin \iota  \cos 2 \omega \sin (\Omega -\Omega _3)-2 \sin 2 \iota \sin 2 \iota _3 \sin 2 \omega \cos (\Omega -\Omega _3)+\sin ^2\iota  (3 \cos 2 \iota _3+1) \sin 2 \omega  \biggr ] ,  
 \nonumber \\
{d \iota\over d \tau}&= {3 \pi  \alpha ^3 m_3 \over {4 \mathcal{A}^3\sqrt{1-e^2} (1-{E}^2)^{3/2}  {m} }}  \[ \sin \iota  \sin \iota _3 \cos (\Omega -\Omega _3)+\cos \iota \cos\iota _3  \] \biggl ( \sin\iota _3 \sin (\Omega -\Omega _3) (5 e^2 \cos 2 \omega+3 e^2+2) 
 \nonumber \\
&+5 e^2 \sin 2 \omega \[ \sin \iota _3 \cos \iota  \cos (\Omega -\Omega _3)-\sin \iota  \cos\iota _3 \] \biggr ),   
\nonumber \\
{d\Omega\over d\tau}&= {3 \pi  \alpha ^3 m_3 \over {4 \mathcal{A}^3 \sqrt{1-e^2}(1-{E}^2)^{3/2}  {m} }} \[ \sin \iota _3 \cos (\Omega -\Omega _3)+\cos \iota _3 \cot \iota \] \biggl ( 5e^2 \sin \iota _3 \sin 2 \omega \sin (\Omega -\Omega _3)   
\nonumber \\
&+(5 e^2 \cos 2 \omega-3 e^2-2) \[ \sin \iota \cos \iota _3-\sin \iota _3 \cos \iota\cos (\Omega -\Omega _3) \]  \biggr),  
 \nonumber \\
{d {\varpi} \over d\tau}&={3 \pi  \alpha ^3 \sqrt{1-e^2} m_3 \over {8 \mathcal{A}^3 (1-{E}^2)^{3/2}  {m} }} 
\biggl (10 \sin \iota \sin 2 \iota _3 \sin 2 \omega \sin (\Omega -\Omega _3) -10 \sin^2 \iota _3 \cos \iota \sin 2 \omega \sin (2\Omega -2\Omega _3)
\nonumber \\
&+\sin^2\iota _3 \cos (2\Omega -2\Omega _3) \[ 2 \sin ^2\iota (4-5 \cos^2 \omega)+20 \cos^2 \omega-10 \] 
\nonumber \\
&+\sin 2 \iota \sin 2 \iota _3  \cos (\Omega -\Omega _3) (3-5 \cos 2 \omega) +(3 \cos 2 \iota _3+1) \[ \sin ^2\iota(5 \cos^2 \omega-4)+1 \] \biggr ), 
\nonumber \\
{d E\over d\tau}&=0, 
\nonumber \\
{d\iota_3\over d\tau}&=-{3 \pi  \alpha ^{7/2} m_1 m_2 \sqrt{M} \over {4 \mathcal{A}^{7/2} (1-{E}^2)^2 {m} ^{5/2}}} \biggl ( \cos \iota _3 \biggr \{ \sin 2 \iota \sin (\Omega -\Omega _3) (-5 e^2 \cos ^2\omega+4 e^2+1)-5e^2 \sin \iota  \sin 2 \omega  \cos (\Omega -\Omega _3) \biggr \} 
\nonumber \\
&+\sin \iota _3 \biggl \{ \sin (2 \Omega -2 \Omega _3) \[ \sin ^2\iota  (-5 e^2 \cos ^2\omega+4 e^2+1)+10 e^2 \cos ^2\omega-5 e^2 \]
+5 e^2 \cos \iota  \sin 2 \omega  \cos (2 \Omega -2 \Omega_3) \biggr\}  \biggr ), 
\nonumber \\
{d{{\Omega}}_3\over d\tau}&= -{3 \pi  \alpha ^{7/2} m_1 m_2 \sqrt{M} \csc\iota _3 \over {8 \mathcal{A}^{7/2} (1-{E}^2)^2  {m} ^{5/2}}} \biggl( \cos 2\iota _3 \biggl \{ \sin 2 \iota \cos (\Omega -\Omega _3) (5 e^2 \cos 2 \omega -3 e^2-2)-10 e^2 \sin \iota  \sin 2\omega \sin (\Omega -\Omega _3)  \biggr \} 
\nonumber \\
&+\sin 2 \iota _3 \biggl \{ \frac{1}{2} \sin ^2\iota \[ \cos ( 2\Omega -2 \Omega _3)+3 \] (5 e^2 \cos 2 \omega -3 e^2-2)  +5 e^2 \cos \iota  \sin 2 \omega  \sin( 2\Omega -2 \Omega _3)
\nonumber \\
&-5 e^2 \cos 2 \omega  \cos ( 2\Omega -2 \Omega _3)+3 e^2+2  \biggr \}
 \biggr ), 
\nonumber \\
{d{\varpi}_3\over d\tau}&={3 \pi  \alpha ^{7/2} {m_1} {m_2} \sqrt{{M}} \over {16 \mathcal{A}^{7/2} (1-{E}^2)^2 {m}^{5/2}}} \biggl ( 30 e^2 \sin \iota\sin 2\iota _3 \sin 2\omega \sin (\Omega -\Omega _3)-30 e^2 \sin ^2\iota _3 \cos \iota \sin 2\omega \sin (2 \Omega -2 \Omega _3) 
\nonumber \\
&+3 \sin ^2\iota _3 \cos (2 \Omega -2 \Omega _3) \[ \sin ^2\iota (-5 e^2 \cos 2\omega+3 e^2+2)+10 e^2 \cos 2\omega \]  
\nonumber \\
&+3 \sin 2\iota \sin 2\iota _3 \cos (\Omega-\Omega _3) (-5 e^2 \cos 2\omega+3 e^2+2)+(2-3 \sin ^2\iota _3) \[ \sin ^2\iota (15 e^2 \cos 2\omega-9e^2-6)+6 e^2+4 \] \biggr ), 
\end{align}
where the time derivation ${d/ dt}$ is  converted to a dimensionless one ${d/d\tau}$ by rescaling time compared to the inner orbital period with $\tau\equiv t/T_{\text{{in}}}={t \over 2\pi}\sqrt{m\over {\alpha}^3}$. 
 
The Kozai-Lidov oscillation of the inner orbit is approximately described by the first four equations in Eq.~(\ref{quadrupolegeneral}), and the rest is the back reaction 
 on the outer orbit which is ignorable. If the outer orbital plane is nearly a constant, then the reference frame could be set approximately on the orbital plane where ${\mathbf{J}}_{\text{out}}$ is depart from $Z$ axis by a very small angle with $\iota_3\to 0$. 
Then the generalized Kozai-Lidov equations in Eq.~(\ref{quadrupolegeneral})  we could be expanded by $\iota_3$ as:
\begin{align} \label{quadrupolestandard}
\frac{de}{d\tau} & = \frac{15 \pi}{2} { \alpha^3 m_3 {e(1-e^2)^{1/2}} \over {\mathcal{A}}^3  m {(1-E^2)^{3/2}} }  \sin^2 (\iota +\iota _3) \sin \omega \cos \omega +O(\iota_3) \,,
\nonumber \\
\frac{d \iota}{d\tau} &= - \frac{15 \pi}{4} { \alpha^3 m_3 e^2 \over  {\mathcal{A}}^3 m {(1-e^2)^{1/2}(1-E^2)^{3/2}} }   \sin 2(\iota +\iota _3)  \sin \omega \cos \omega +O(\iota_3) \,,
\nonumber \\
\frac{d\Omega}{d\tau} &= - \frac{3\pi}{4}  { \alpha^3 m_3 \over  {\mathcal{A}}^3 m {(1-e^2)^{1/2}(1-E^2)^{3/2}} }  \frac{\sin 2(\iota +\iota _3) }{\sin \iota} (1+4e^2 - 5e^2 \cos^2 \omega ) +O(\iota_3) \,,
\nonumber \\
\frac{d\varpi}{d\tau} &= \frac{3\pi}{2}  { \alpha^3 m_3 {(1-e^2)^{1/2}} \over  {\mathcal{A}}^3 m {(1-E^2)^{3/2}} }  \left [ 1- \sin^2 (\iota +\iota _3)  ( 4 - 5 \cos^2 \omega \right ) ] +O(\iota_3) \,.
\end{align}
Which degenerate to the standard Kozai-Liodv formula (see e.g. \cite{Naoz:2016vh, Will:2017vjc}) in the dominant terms, and the dependence on $\Omega-\Omega_3$ is of order $O(\iota_3)$ smaller. 
This means when the outer orbital plane do not change significantly, we could safely use the standard Kozai-Lidov formula to describe the Newtonian quadrupole perturbations. While when the outer orbital plane changes moderately, the secular dynamics will depend on the angle $\Omega-\Omega_3$, and in this case we have to use the Kozai-Lidov formula in the generalized form (\ref{quadrupolegeneral}).  
  \end{widetext}
 
Thought the equations in (\ref{quadrupolegeneral}) seems a bit more complex compared to the standard one, it is a more general description of the Newtonian quadrupole perturbation which could be extended to the case where the outer orbital angular momentum is evolving. And the generalized Kozai-Lidov formula in Eq.~(\ref{quadrupolegeneral}) will certainly degenerate to the standard Kozai-Lidov oscillation dynamically with an approximately constant outer orbital plane, as it is guaranteed by the equations of motion. 
This degeneration happens in two situations. On the one hand, up to Newtonian order, the total orbital angular momentum of the three body system is strictly conserved. This conservation could simplify the formula in Eq. (\ref{quadrupolegeneral}) by the fact that the relation $\Omega-\Omega_3=\pi$ is precisely granted \cite{Will:2017vjc}. On the other hand, if the other orders of perturbations do not change the outer orbital plane, like the de-Sitter precession (as shows in the next part of this section), the dynamics on $\Omega-\Omega_3$ is decoupled with the Newtonian quadrupole perturbations. An analogous discussion also suits the other orders of Newtonian perturbations, like the octupole \cite{Naoz2011Nature} and the hexadecapole order perturbations \cite{Will:2017vjc}.

The binary 1PN acceleration in Eqs.~(\ref{abinary}) and (\ref{Abinary}) induce the typical relativity precession on the pericenter $\omega$ ($\omega_3$) of the binary system by
 \m
 {d \omega\over d\tau}&=&{6\pi m\over p}, \ 
  {d \omega_3 \over d\tau}={T_{\text{in}}\over T_{\text{out}}}{6\pi M\over P}.
  \label{abinaryprecession}
 \n

In this paper, we derive the secular evolutions contributed by the leading three body 1PN accelerations in our system as presented in Eqs.~(\ref{a3body1PN}) and (\ref{A3body1PN}), the results are listed as bellow, 

 \medskip
 \noindent
{\bf Leading three body 1PN effects}
\begin{align}
  {de \over dt}&={dp\over dt}=0,  \nonumber\\
  {d\iota\over dt}&=-\frac{3 m_3^{3/2}} {2 \mathcal{A}^{5/2} (1-E^2) }  \sin \iota _3 \sin \left(\Omega -\Omega _3\right),   \nonumber\\
  {d\Omega \over dt}&=\frac{3 m_3^{3/2}}{2 \mathcal{A}^{5/2} \left(1- {E}^2\right)}  \[ \cos \iota _3 -\sin \iota _3 \cot \iota  \cos \left(\Omega -\Omega _3\right) \] ,   \nonumber\\
  {d\omega\over dt}&= \frac{3 m_3^{3/2}} {2 \mathcal{A}^{5/2} (1-E^2) }  \sin \iota _3 \csc \iota  \cos \left(\Omega -\Omega _3\right) ,  \nonumber\\
  {dE\over dt}&= {dP\over dt}={d{\iota_3}\over dt}={d{\Omega_3}\over dt}={d{\omega_3}\over dt}=0 , 
  \label{threebodyprecessiongeneral}
\end{align}
the precession on $\iota$ and $\Omega$ in Eq.~(\ref{threebodyprecessiongeneral}) gives the de-Sitter precession of ${\mathbf{J}}_{\text{in}}$ about ${\mathbf{J}}_{\text{out}}$ and the precession on $\omega$ gives the precession of the Runge-Lenz vector ${\bf{Q}}$ about ${\mathbf{J}}_{\text{out}}$. These precessions together describe the de-Sitter precession in a general reference frame where the $Z$ axis is not necessarily set to the outer orbit angular momentum. But when $Z$ is aligned with ${\mathbf{J}}_{\text{out}}$ and $\iota_3 \to 0$, only $\Omega$ is affected in the de-Sitter effect as is described bellow  
\begin{align}
  {d\Omega \over dt}&=\frac{3 m_3^{3/2}}{2 \mathcal{A}^{5/2} \left(1- {E}^2\right)}  +O(\iota_3),   \nonumber\\
   {d\iota \over dt}&=O(\iota_3),  {d\omega\over dt}=O(\iota_3) . 
  \label{desitterprecession}
\end{align}

For an evolving outer orbital plane, the de-Sitter precession couples in the secular evolutions through the generalized Kozai-Lidov formula. And since the third body is a SMBH in our triple system, the de-Sitter precession could be the same order or even larger than the (inner) binary 1PN precession under a large parameter space. Thus, the three body 1PN effects dominated by the de-Sitter precession could couple in the secular dynamics at zero order in the multiple-scale analysis method (which is defined in such as Eq.~(12.235) of \cite{Poisson:2014aa}, or Eq.~(3.50) of \cite{Lim:2020cvm}).

The secular dynamics from the spin effects of the SMBH is resulted by accelerations in (\ref{inner_spin_acceleration}) and (\ref{outer_spin_acceleration}), which are calculated in \cite{Fang:2019hir}. The non-vanishing results are, 

\medskip
\noindent
{\bf spin effects at 1.5PN order}
 \begin{align}
{d \iota\over d t}=&{3  a  {m_3} \over 4 \mathcal{A}^3 (1-{E}^2)^{3/2} } \sin 2 \iota _3 \sin (\Omega -\Omega _3), \nonumber\\
{d\Omega\over d t}=& -{  a {m_3}  \over 4 \mathcal{A}^3 (1-{E}^2)^{3/2} } \nonumber\\
&\times \[ -3 \sin2 \iota _3 \cot\iota \cos (\Omega -\Omega _3)+3 \cos2 \iota _3+1 \], \nonumber\\
{d\omega\over d t}=& -{3  a  {m_3}  \over 4 \mathcal{A}^3 (1-{E}^2)^{3/2} } \sin2 \iota _3 \csc\iota \cos (\Omega-\Omega _3), \nonumber\\
{d\Omega_3\over d t}=&{2  a  {m_3}\over \mathcal{A}^3 (1-{E}^2)^{3/2} }, \nonumber\\
{d \omega_3\over d t}=&-{6  a  {m_3} \over \mathcal{A}^3 (1-{E}^2)^{3/2} } \cos\iota _3, 
\label{leadingspin}
  \end{align}
where the change of $\Omega_3$ in Eq.~(\ref{leadingspin}) is the Lense-Thirring precession on the outer orbit. The inner orbit is not simply precessing around the spin axis, but in a rather complex way. 
Note that up to the orders considered in our full text, the outer orbital pericenter $\omega_3$ is decoupled from the evolution of other orbital elements. The coupling of $\omega_3$ happens at octuple order which is usually small due to the small ratio of $\alpha/\mathcal{A}$, so it is ignored in this context, though, for a full parameter space of evolution, it is needed to consider. The triple systems where the octuple effect matters is studied such as in \cite{Lithwick:2011hh}.

\section{Numerical results}
\label{numericalresult}

\subsection{On the dynamical evolution}
\label{dynamicalevolution}
In this subsection, we display the numerical results of our SMBH-BBH three body system. 
We begin with a group of initial data with a BBH of mass $m_1=m_2=20\Msun$ (or $30\Msun$), which are the typical mass detected by LIGO/Virgo. The third body is a SMBH with $m_3=4\times10^6\Msun$ which is similar to the one in our galaxy center. The inner binary is separated with semi-major axis $\alpha=0.04$ AU and has an initial eccentricity $e=0.1$. They are set to a distance of $\mathcal{A}=30$ AU to the SMBH with an outer orbit eccentricity $E=0.1$. The line of apsides of the two orbits are set to $X$ axis thus $\Omega=\Omega_3=\omega= \omega_3=0$, and the inclination angle between the two orbits is simply $\iota-\iota_3$ for convenience but not lose generality. 

\begin{widetext}
\begin{figure*}[htp] 
\centering
\includegraphics[width=1 \linewidth]{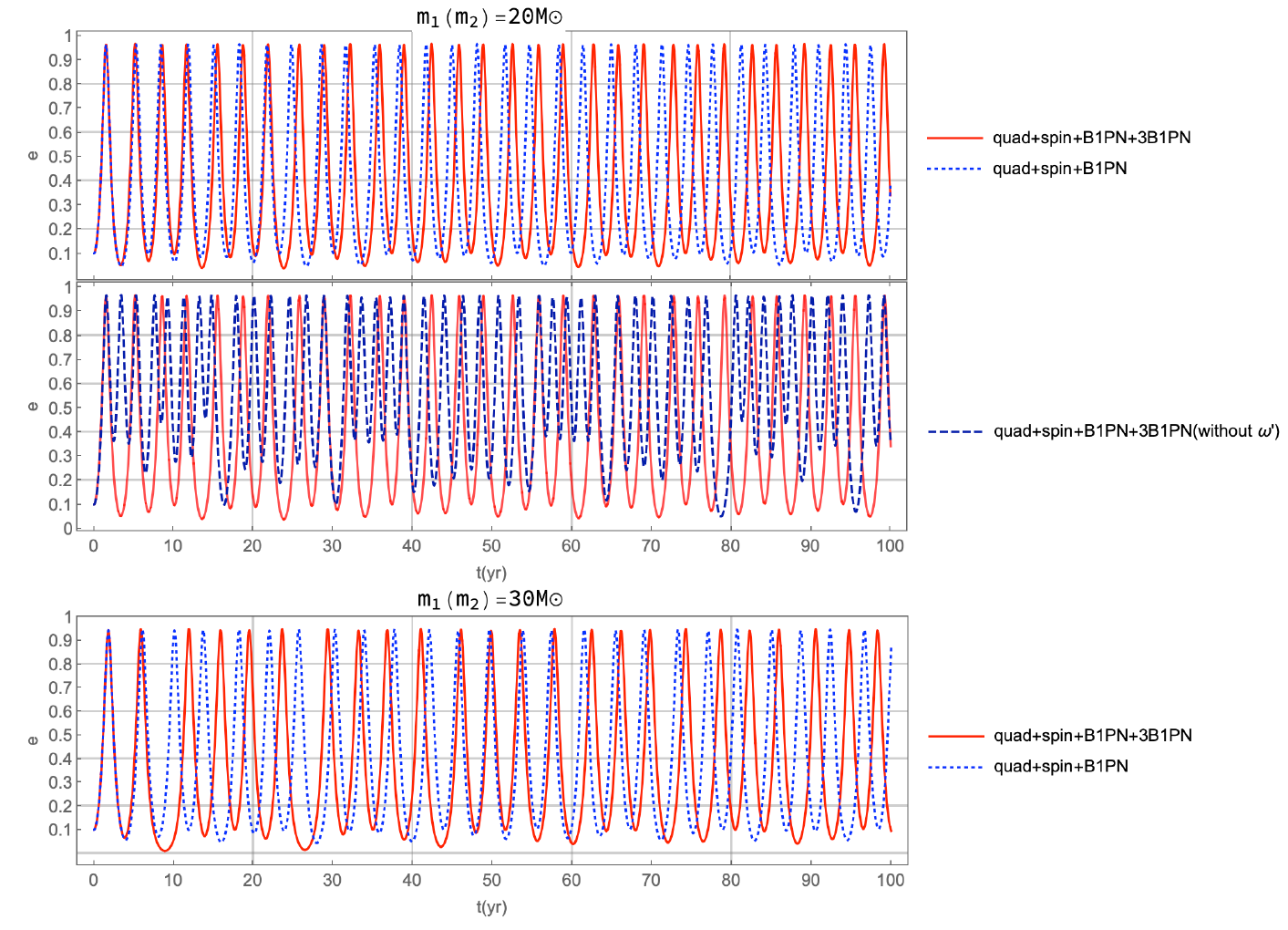} 
\caption{The evolution of eccentricity in 100 years. The initial conditions for the two panels are: $\iota_3=60^{\circ}$ and $\iota=140^{\circ}$, thus the initial angle between $\mathbf{J}_{\text{in}}$ and $\mathbf{J}_{\text{out}}$ is $\iota-\iota_3=80^{\circ}$, and the SMBH has a spin parameter of $a=0.9m_3$. In the upper panel we have a BBH of mass $m_1=m_2=20\Msun$, while the lower line have a BBH with $m_1=m_2=30\Msun$. 
In the red line, we have included all the dynamical effects considered in this work, including the Newtonian quadrupole effect (Kozai-Lidov) in Eq.~(\ref{quadrupolegeneral}), the binary 1PN precession in Eq.~(\ref{abinaryprecession}), the three body 1PN effects in Eq.~(\ref{threebodyprecessiongeneral}), the spin effects in Eq.~(\ref{leadingspin}), and radiation reaction \cite{Peters:1963ux} (this effect hardly contribute within 100 years here, but it matters for a life-time evolution). The dotted blue line has included the same dynamical effects considered in the red line except the three body 1PN effects in Eq.~(\ref{threebodyprecessiongeneral}), and the dashed dark bule line only lacks the precession effect on $\omega$ in Eq.~(\ref{threebodyprecessiongeneral}) compared to the red line.   
}
\label{e_evolutions}
\end{figure*} 
\end{widetext}

In Fig.~\ref{e_evolutions}, we display the numerical evolution of eccentricity. In the upper panel, we choose the BBH with an equal mass of $20\Msun$, and compare the results of evolution with all the dynamics considered in this work with the results when the three body 1PN effects (de-Sitter precession) or the precession on $\omega$ in the three body 1PN effects are absent. We could see that neither the absent of the de-Sitter precession nor the absent of the precession on $\omega$ in de-Sitter effects could lead to correct evolutions. 
This indicates the de-Sitter precession is needed to be considered when there is a SMBH spin effect, which include both the precession of the inner orbital angular momentum and the Runge-Lenz vector around the outer orbit angular momentum. 
In the lower panel, we choose a BBH with mass $30\Msun$ while keep the other initial conditions and line styles the same to the upper panel. In the second case, the difference of the results due to the absent the three body 1PN effects is more obvious.

As a comparison, it is trivial to verify that the de-Sitter precession decouples when the Lense-Thirring precession disappears when $a=0$ or initially $\iota_3=0$. 
These indicate the three body 1PN effects (which is dominated by the de-Sitter precession) is coupled in the secular dynamics without spin effects or the outer orbital Lense-Thirring precession. 

\subsection{On the gravitational waves}
\label{onGW}
In the final, we will move onto the study of the impact of the spin effects combined with the de-Sitter precession on the GW singles of the BBH. The maximal eccentricity of the inner orbit excited by the Kozai-Lidov oscillation is found limited to the vertical inner orbit angular momentum (which is proportional to $\Theta=\sqrt{1-e^2}{{\mathbf{h}}\cdot {\mathbf{H}}}$) in the test particle approximation by \cite{Lithwick:2011hh, Naoz:2016vh}:
\m
e_{\text{max}}=\sqrt{1-{5\over 3}{\Theta}^2} , 
\label{emax}
\n
 which is calculated with a small value for eccentricity and zero for $\omega$ initially.   
Eq.~(\ref{emax}) also holds in our triple system where $m_3\gg m_1, m_2$. Since in this case, we have $\mathbf{J}_{\text{out}} \gg \mathbf{J}_{\text{in}}$, thus $\mathbf{J}_{\text{out}}$ is not affected by inner orbital motion, then the projection of $\mathbf{J}_{\text{in}}$ along $\mathbf{J}_{\text{out}}$ is conserved at the Newtonian quadrupole order. Combining this with the Hamiltonian beyond the test particle limit (see e.g. Eq.~(22) of \cite{Naoz:2016vh}) will lead to the result in Eq.~(\ref{emax}). 
And since the de-Sitter precession do not change the value of $\Theta$,  here $\Theta$ changes only due to the spin of the SMBH by
\m
\Theta'(t)={3 \sqrt{1-{e}^2} ({\bf{S}}\times {{\bf{H}}}) \cdot {{\bf{h}}}  \over 2 \mathcal{A}^3 (1-E^2)^{3/2} },  
\label{spin_elements_general}
\n
where $\bf{S}$ is the spin angular momentum, $\bf{h}$ and $\bf{H}$ are the unit direction of $\mathbf{J}_{\text{in}}$ and $\mathbf{J}_{\text{out}}$ respectively as discussed before. 

\begin{figure}[h] 
\centering
\includegraphics[width=1 \linewidth]{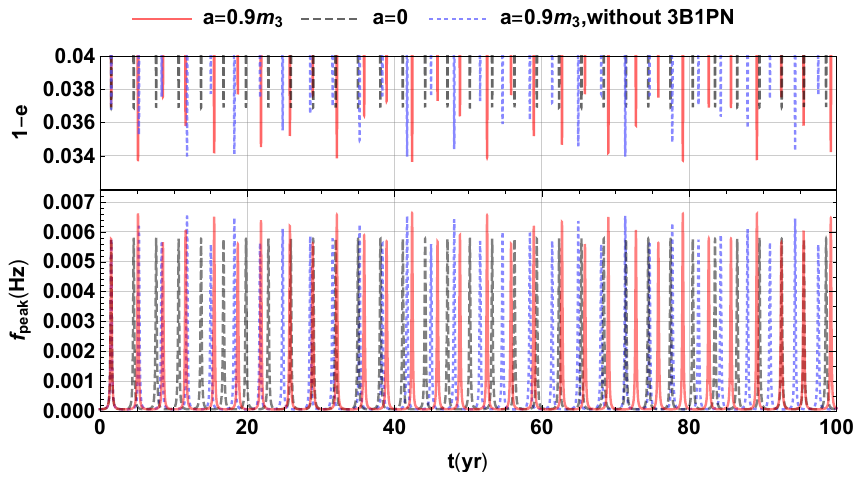} 
\caption{Upper panel: zoom in of the eccentricity evolution in the upper panel of Fig.~\ref{e_evolutions} near its maximal values.  
Lower panel: the peak frequency of the GWs correspond to the eccentricity in the upper panel. 
The red solid line and the blue dotted line have the same line styles with that in Fig.~\ref{e_evolutions}, while the dashed black line is only different from the red line by the spin parameter of the SMBH with $a=0$. 
}
\label{emax_fpeak_t}
\end{figure}

The maximal eccentricity excited by the Kozai-Lidov oscillation is modulated by the spin of the SMBH through Eq.~(\ref{emax}) and (\ref{spin_elements_general}). 
 And the peak frequency of the GW is closely related to the eccentricity by $f_{\text{peak}}={\sqrt{m}(1+e)^{-0.3046} \over \pi \left[\alpha(1-e)\right]^{3/2}} $\cite{Wen:2002km}. In Fig.~\ref{emax_fpeak_t}, we illustrate the evolutionary behaviors of the maximal values of the eccentricity in the upper panel (and thus $f_{\text{peak}}$ in the lower panel) which modulated by spin (red solid line), and also the results when either spin is zero (black dashed line) or the de-Sitter effect is absent (blue dotted line) as a comparison.    
In this example, the spin effects will cause the closest distance between the BBH $(1-e_{\text{max}})\alpha$ to change by several percent in the neighboring Kozai-Lidov circles. As a result, the corresponding maximal values of $f_{\text{peak}}$ could reach to a difference of nearly $0.001$Hz as shows in the lower panel. 
The ignoring of the de-Sitter precession when spin is $a=0.9 m_3$ will lead to a shift in the phase and a small change in the amplitude of the Kozai-Lidov oscillation. 
These maximal values remain constant within several Kozai-Lidov circles without spin while they change due to spin effects, these are the unique characteristic resulted from spin here.   
Interestingly, when the eccentricity of the BBH is excited to a relatively large number, as shows in this case, the peak frequency of the GWs could locate in LISA band. 

The detectability of the Kozai-Lidov oscillation of the BBH near a (non-spinning) SMBH due to their GWs by the detection of  LISA has been studied in \cite{Hoang:2019kye}. 
And the detectability of the spin effects from the SMBH through these GW sources have been studied in \cite{Fang_2019} which compared the fitting factor of the GWs calculated with a SMBH of spin 0.9 and spin zero within four years, while the de-Sitter precession is not considered in this article. 
Here, we step the topic on probing the spin of SMBH a little further by looking at the unique characteristics due to the spin effects shows in Fig.~\ref{emax_fpeak_t}. 

To see if we could discriminate between the cases with and without SMBH spin effect, or with and without the de-Sitter precession effect from only one or two GW peaks in Fig.~\ref{emax_fpeak_t} by LISA mission, we calculate the single to noise ratio (SNR) of the GWs which is defined for example in Eq.~(55) of \cite{Barack:2003fp}. In our case, the semi-major axis and the Kepler orbital frequency is approximate a constant within the first several years, the SNR in the frequency domain could be converted to the time domain by Parseval’s theorem as \cite{Barack:2003fp}
\m
\langle SNR^2 \rangle=\sum_{n=1}^{\infty}{2 \over 5 \pi^2 D^2} {1\over {f_n}^2 S_{h}(f_n)} \int  {\dot{E}}_{n} d t, 
\label{SNRGWPower}
\n
where ${\dot{E}}_{n}$ is the GW radiation power given by
\e
{\dot{E}}_{n} ={32\over 5}\mu^2 m^{4/3}(2\pi f)^{10/3}g(n,e) , 
\q
where $\mu={m_1 m_2\over m}$, and 
\m
g(n,e)&=&{n^4\over 32}\{[J_{n-2}(ne)-2e J_{n-1}(ne)+{2\over n}J_{n}(ne) \nonumber\\
&&+2e J_{n+1}(ne)-J_{n+2}(ne) ]^2           \nonumber\\
&&+(1-e^2)[J_{n-2}(ne)-2J_{n}(ne)+J_{n+2}(ne)]^2    \nonumber\\
&&+ {4\over 3 n^2}[J_{n}(ne)]^2 \}.
\n

When assuming the source in our galaxy center, the SNR for the first two peaks in Fig.~\ref{emax_fpeak_t} with spin $a=0.9 m_3$ are respectively 260 and 283, while the SNR for the first two peaks with spin zero are both 260. These results are consistent with the evolution of the maximal eccentricity: the larger the eccentricity, the larger of the radiation power of GWs. 
The SNR difference $\Delta \text{SNR}(\Delta e_{\text{max}})$ between any two peaks in the red line varies from peak to peak, but can at times reach up to 23, such as the first two. 
 The separation between the nearby two peaks is about four years, thus is potentially detectable by a four-year LISA mission especially when LISA is extended to ten years.  Besides, the SNR for the first two peaks when spin is $a=0.9 m_3$ while the de-Sitter effect is absent are respectively 261 and 271, this indicates that considering the 3B1PN effect is necessary when modeling the GW signal. 


\section{Conclusion}
\label{conclusion}

In this paper, we study the secular evolution of the SMBH-BBH system up to 1.5PN order where the SMBH have a large spin.  
We resolving the three-body 1PN effects starting from the Einstein-Infeld-Hoffmann equations of motion \cite{Einstein1938}, which the dominant effect is the de-Sitter precession on the inner orbit. We conduct the double average of the Lagrange planetary equations to get the secular evolutionary equations with the help of Mathematica software. 

The de-Sitter precession is previously considered as a sub-leading effect in the three body systems either with a small third body \cite{Will:2013cza, WillPRL2014} or in a case when it is decoupled in the zero order secular equations \cite{Lim:2020cvm}. While in our triple system, the large mass of the SMBH could cause the de-Sitter precession to reach to or even larger than the amplitude of the binary 1PN effect in a large parameter space. More over, the spin effects from the SMBH will cause the outer orbital plane to precess due to the Lense-Thirring effect, 
 the evolving outer orbital plane causes the generalized Kozai-Lidov effect to depend on the angle between the two lines of nodes $\Omega-\Omega_3$ \cite{Fang:2019hir}, thus couples the de-Sitter precession in the secular dynamics and contribute significantly in the evolutions. 
We state that the de-Sitter precession include both  the precession of the inner orbital angular momentum and its Runge-Lenz vector around the outer orbit angular momentum in a general reference frame where the Z axis is not set to the outer orbit angular momentum due to the Lense-Thirring precession. 
This is different from the description of the de-Sitter precession only with a precession of the inner orbital longitude of ascending node when seeing from a reference frame where the outer orbit angular momentum is alone $Z$ axis.
Our general argument on the coupling of the three body 1PN effects could be extended to any situation where the outer orbital plane is evolving due to other mechanisms, such as a non-spherical gravitational potential \cite{Ivanov10.1111, Merritt2011ApJ}. 

We show numerical results of the impact of spin effect and the de-Sitter effect on the evolution and the GW singles of the BBH. 
The spin effect from the SMBH modulates the Kozai-Lidov oscillation both in the phase and the amplitude \cite{Fang:2019hir, Fang_2019}, and the maximal eccentricity excited by the Kozai-Lidov oscillation is evolving at Kozai-Lidov timescale due to spin, which is caused by spin uniquely in our consideration. 
We also show that without the de-Sitter precession effect, the modulation behavior of the spin effect will be different. 
Our numerical result indicates that the spin effect is detectable by two nearby GW peaks in our representative example, and the de-Sitter effect is not ignorable when calculating the SNR and analysis the spin effect. These results could lead to a potential way to prob the SMBH spin effect or the spin parameter the SMBH by LISA in the future, in which the de-Sitter precession effect is needed to be considered.

Acknowledgments. We would like to thank Xian Chen and Misao Sasaki for useful discussions. 
This work is supported by grants from NSFC (grant No. 11975019, 11690021, 11991053, 11947302), the Strategic Priority Research Program of Chinese Academy of Sciences (Grant No. XDB23000000, XDA15020701), and Key Research Program of Frontier Sciences, CAS, Grant NO. ZDBS-LY-7009.
	
	


	
%

\end{document}